\documentclass[a4paper, titlepage]{article}    
\usepackage{authblk}
\usepackage[usenames,dvipsnames,svgnames,table]{xcolor}
\usepackage{amsmath}
\usepackage{hyperref}

\newcommand{\tool}[1]{\textbf{#1}}
\newcommand{\Rcode}[1]{\texttt{#1}}

\usepackage[pdftex]{graphicx}
\usepackage{float}
\graphicspath{{./Pictures/}}
\newfloat{suppfig}{tbh}{lop}
\floatname{suppfig}{Supplementary Figure}
\usepackage{caption}
\usepackage{listings}
\renewcommand{\thefootnote}{\fnsymbol{footnote}}

\begin{document}

\title{Robustly detecting differential expression in RNA sequencing data using observation weights}

\author[1,2]{Xiaobei Zhou}
\author[1,2]{Helen Lindsay}
\author[1, 2,  \footnote{To whom correspondence should be addressed.
Tel: +41 44 635 48 48; Fax: +41 44 635 68 68; Email: mark.robinson@imls.uzh.ch} ]{Mark D. Robinson }

\affil[$^{1}$]{Institute of Molecular Life Sciences, University of Zurich, CH-8057 Zurich, Switzerland}
\affil[$^{2}$]{SIB Swiss Institute of Bioinformatics, University of Zurich, CH-8057 Zurich, Switzerland}

\maketitle

\renewcommand{\thefootnote}{\arabic{footnote}}\setcounter{footnote}{0}

\begin{abstract}
A popular approach for comparing gene expression levels between (replicated) conditions of RNA sequencing data relies on counting reads that map to features of interest.  Within such count-based methods, many flexible and advanced statistical approaches now exist and offer the ability to adjust for covariates (e.g., batch effects).  Often, these methods include some sort of {\em sharing of information} across features to improve inferences in small samples.  It is important to achieve an appropriate tradeoff between statistical power and protection against outliers.  Here, we study the robustness of existing approaches for count-based differential expression analysis and propose a new strategy based on observation weights that can be used within existing frameworks.  The results suggest that outliers can have a global effect on differential analyses.  We demonstrate the effectiveness of our new approach with real data and simulated data that reflects properties of real datasets (e.g., dispersion-mean trend) and develop an extensible framework for comprehensive testing of current and future methods.  In addition, we explore the origin of such outliers, in some cases highlighting additional biological or technical factors within the experiment.  Further details can be downloaded from the project website: \\ {\scriptsize \url{http://imlspenticton.uzh.ch/robinson_lab/edgeR_robust/}}
\end{abstract}

\section{Introduction}
RNA sequencing (RNA-seq) is widely used for numerous biological applications, including the detection of alternative splice forms, RNA editing, allele-specific expression profiling, novel transcript discovery but most notably, for detecting changes in expression between experimental conditions or treatments. Compared to  microarray technology, RNA-seq offers an open system, higher resolution, lower relative cost and less bias \cite{Wang2009a}. A typical RNA-seq experiment includes: i) capture of an RNA subpopulation (e.g., polyA-enriched, depletion of ribosomal RNA) from cells of interest; ii) reverse transcription into complementary DNA (cDNA); iii) preparation and sequencing of millions of short cDNA fragments ($\sim$200bp); iv) mapping to a reference genome or (assembled) transcriptome; v) counting according to a catalog of features.  This last counting step can be conducted by excluding ambiguous reads between genes \cite{Anders2013}, or with advanced tools that portion ambiguous reads to transcripts \cite{Li2011} or can be done in combination with assembly tools \cite{Grabherr2011}.  The focus here is on methods for count-based differential expression (DE) analyses and the robustness thereof; thus, the starting point here is a count table of features-by-samples, such as those available from the \tool{ReCount} project \cite{Frazee2011}.

\begin{figure*}[ht]                 
\centering
\includegraphics[width=\textwidth]{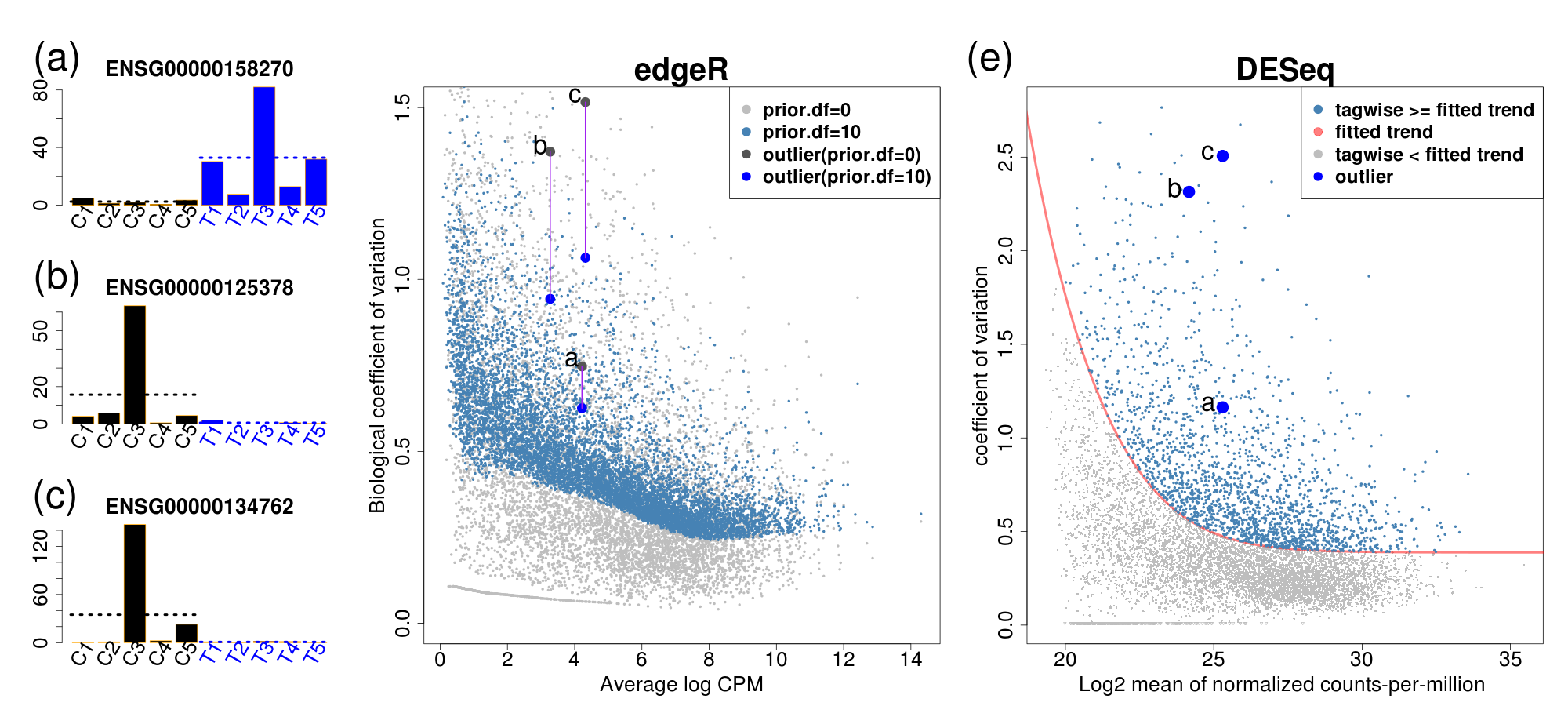}
\caption{From Pickrell \cite{Montgomery2010} data ten randomly selected samples from individuals  are divided into two groups of 5, forming an artificial ``null'' scenario. (a), (b) and (c) show barplots of three genes from top 10 DE genes with one or two extremely large observations. (d) and (e) plot genewise biological coefficient of variation (BCV) against gene abundance (in log2 counts per million) for \tool{edgeR} and \tool{DESeq}. In panel (d), gray dots shows unmoderated biological BCV estimates ($\sqrt{\phi_i}$) (prior degrees of freedom$=0$). Steel blue dots shows moderated biological BCV with prior degree $10$ (default setting for \tool{edgeR}). Three outlier genes on (a), (b) and (c) are labelled by large blue dots. For (e) \tool{DESeq} uses  the maximum (steel blue dots) of a fitted dispersion-mean trend (red line) or the individual feature-wise (tagwise) dispersion estimate. Three outlier genes are also pointed out by large blue dots.} 
\label{fig:illustration}
\end{figure*}

Considerable recent effort has been paid by the statistical community to the discovery of DE features, given a count table; recent comparisons have shown that no method dominates the spectrum of possible situations \cite{Soneson2013, 2013arXiv1301.5277R}.  RNA-seq remains expensive and in many cases researchers are studying precious samples or rare cell types, so the number of biological replicates is often limiting.  It is clear that the most successful methods implement some form of {\em information sharing} across the whole dataset to improve DE inference \cite{Anders2013}, and this becomes an intricate exercise to tradeoff power, false discovery control and protection against outliers.  To highlight this distinction, we describe two popular software implementations for the negative binomial (NB) model, which arguably is the {\em de facto} standard for accounting for biological variability in such genome-scale count datasets.  The latest version of \tool{edgeR} moderates dispersion estimates towards a trended-by-mean estimate \cite{McCarthy2012}, whereas \tool{DESeq} takes the maximum of a fitted dispersion-mean trend or the individual feature-wise dispersion estimate\cite{Anders2010}.  The effect imposed on features with ``outliers'' is illustrated in Figure \ref{fig:illustration}.  Ten randomly selected samples from individuals from the HapMap project (denoted as Pickrell \cite{Montgomery2010}) are divided into two groups of 5, forming an artificial ``null'' scenario.  While very little true differential expression is expected, a low rate of false detections occur; in particular, \tool{edgeR} detects a small number of genes with low estimated false discovery rate that exhibit one or two observations that are generally much higher in expression (Figure \ref{fig:illustration}a-c).  We believe that there are two causes for this: i) the sensitivity of relative expression estimates to these ``outlying'' observations; ii) moderation of the dispersion estimates toward the trend.  In contrast, \tool{DESeq} remains largely unaffected by these outliers, since the dispersion estimation policy is to keep the maximum; in what follows, we will explore effect of this maximum policy on power.  All computed statistics for this dataset are stored in Supplementary Table \textbf{1}.

\enlargethispage{-65.1pt}

The downstream effect of these dispersion estimation strategies suggest: i) \tool{DESeq} is conservative but robust; ii) \tool{edgeR} can be sensitive to outliers when there is sufficient dispersion smoothing towards the trend (effectively underestimating the dispersion in the shrinking process), but should be more powerful in the absence of such extreme observations \cite{Anders2013}.  Our goal in the current study is to achieve a suitable middle ground, perhaps forfeiting a small amount in statistical efficiency, similar to established robustness frameworks, to reduce the influence of extreme observations in differential expression calls.  As hinted above and in general, robustness is not solely determined by the dispersion parameter, but also by controlling the influence of outliers to other parameters in the model (e.g., those representing changes in expression).  We explore these aspects in both simulated and real data, provide a  extensible framework for evaluating the tradeoffs and highlight some instances of biology or technical factors that may give outliers.

The literature is rich in alternatives for count-based DE analyses and in particular, dispersion estimation, yet it remains increasingly difficult to assess the performance across the range of  possibilities.  For example, recent evidence suggests that one can suitably transform count data and analyze with methods developed for microarrays, with special treatment \cite{Law}.  The mainstream strategy is to directly fit count data to extensions of the Poisson model and in particular, the NB model.  Many implementations are available as \tool{R}/\tool{Bioconductor} packages \cite{Gentleman2004}, such as \tool{edgeR} \cite{Robinson2010e}, \tool{DESeq} \cite{Anders2010}, \tool{ShrinkBayes} \cite{VanDeWiel2013}, \tool{baySeq} \cite{Hardcastle2011} and variations of dispersion estimation that can be used within existing implementations \cite{Wu01042013}; the main differences lie in the estimation of the dispersion or in the inference machinery (e.g., Bayesian versus frequentist).  Recent comparisons and summaries of the methods available can be found in \cite{Anders2013}, \cite{Soneson2013} and \cite{2013arXiv1301.5277R}.

Some early and existing count-based DE analysis tools only allowed 2-group comparisons.  That is, they could not handle more complex situations, such as paired samples, time courses or batch effects. Recently, McCarthy et al. developed generalized linear model (GLM) capabilities in \tool{edgeR} \cite{McCarthy2012}, allowing a much broader class of experimental designs to be analyzed and other frameworks have followed suit.  However, GLMs require iterative fitting and more complicated dispersion estimation machinery \cite{McCarthy2012}. As shown in Figure \ref{fig:illustration}, this framework can suffer a lack of robustness, whereby even a single extreme value (outlier) could largely affect estimates of regression parameters (e.g., mean of experimental condition), as highlighted by recent comparative studies \cite{Soneson2013} (see also Figure \ref{fig:illustration}).  In addition, the moderation of the dispersion parameter towards a trended value is actually contributing to the lack of robustness, forcing the dispersion to be underestimated \ref{fig:illustration}).  Although the details are not yet published, \tool{DESeq2} (successor of \tool{DESeq}) takes an altogether different stance on robustness: using a Cook's distance metric,  features that exhibit an extreme value are not considered for downstream statistical testing.  

\begin{figure}[ht]                 
\centering
\includegraphics[width=0.5\textwidth]{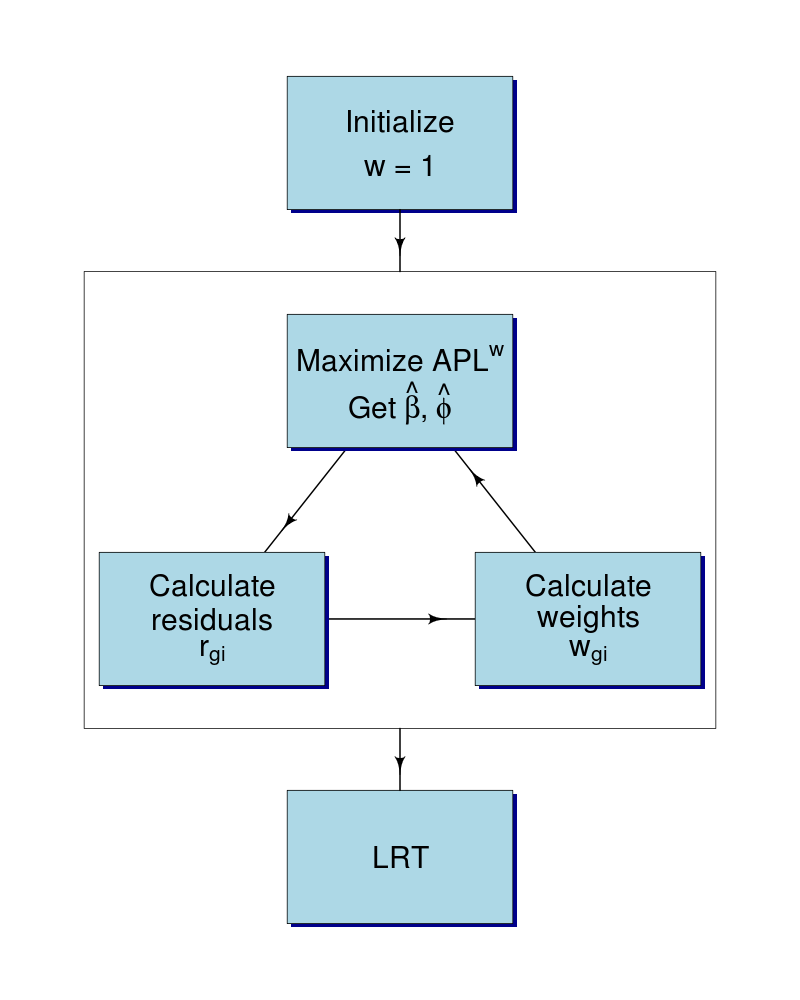}
\caption{The flow chart of the robust algorithm implemented in \tool{edgeR}. $\hat{\beta}$ is the estimated GLM regression coefficient and $\hat{\phi}$ is the moderated dispersion estimate by maximizing APL$^w$ (Equation~\ref{eq:aplw}). $r_{gi}$ is the Pearson residual corresponding to count $y_{gi}$ from Equation~\ref{eq:residual}. $w_{gi}$ is the observation weight from Equation~\ref{eq:weight}. LRT (\Rcode{glmLRT} in \tool{edgeR}) computes likelihood ratio tests.} 
\label{flow}
\end{figure}

The strategy proposed in this paper is that of {\em observation weights}, effectively downweighting outliers to dampen their influence.  There is already some precedent for doing this in GLM settings: Carroll and Pederson \cite{Carroll1993} introduced weighted maximum likelihood estimators for the logistic model; Cantoni \cite{Cantoni2001} presented a robust quasi-likelihood approach for inference in binomial and Poisson models; Agostinelli and Alqallaf \cite{agostinelli_robust_2009} derived weighted likelihood equation for GLMs by directly inserting {\em observation weights} into iterative re-weighted least squares algorithm (IRLS).  Of particular importance, after adding observation weights, the asymptotic theory suggests that likelihood ratio statistics of model parameters still converge to approximate chi-squared distributions under the null hypothesis \cite{Agostinelli2002}.  At present, no ``off-the-shelf'' robust approach is readily available for the negative binomial model in the context of genome-scale computations. In this paper, we build an outlier-resistant framework that maintains high power and achieves decent false discovery control and make it available in the \tool{edgeR} software package; the same strategy could be employed in other frameworks.  We benchmark its performance on real and simulated data and explore the origins of outlying observations. 

\section{MATERIALS AND METHODS}
\subsection{A standard setup of NB model in GLM framework} 

To most easily explain the addition of observation weights, we follow closely the notation used in McCarthy et al. \cite{McCarthy2012}.  Let the $Y_{gi}$ be the read count in sample $i$ for feature $g$. Assume $Y_{gi}$ follows a NB distribution with mean $\mu_{gi}$ and dispersion $\phi_g$, denoted by $Y_{gi} \sim NB(\mu_{gi}, \phi_g)$.  Feature $g$'s variance equals $\mu_{gi} + \phi_g \cdot {\mu_{gi}}^2$, while the dispersion $\phi_g$ represents the square of the {\em biological coefficient of variation} \cite{McCarthy2012}.  In the GLM setting, the mean response, $\mu_{gi}$, is linked to a linear predictor, here with the canonical logarithm link according to:
\begin{equation}
log(\mu_{gi}) = {X} \beta_g + \log N_i, 
\end{equation}
where $X$ is the design matrix containing the covariates (e.g., experimental conditions, batch effects, etc.), $\beta_g$ is a vector of regression parameters (a subset of which are of interest for differential expression inference) and $N_i$ is the (effective) library size for sample $i$.

For estimation of the regression parameters, maximum likelihood estimation is used.  The derivative of the log-likelihood, $l(\beta_g)$, with respect to the coefficient $\beta_g$ is $X^T z_g$, where $z_{gi} = (y_{gi}-\mu_{gi})/(1+\phi_g\mu_{gi})$. The estimated value of $\beta_g$ can be obtained by the iteratively re-weighted least squares algorithm (IRLS) in the form:
\begin{equation}
\label{eq:beta}
\beta_g^\text{new} = \beta_g^\text{old} + (X^T \Omega_g X)^{-1}{X}^Tz_g,
\end{equation}
where $X^T \Omega_g X$ is the Fisher information matrix (also denoted below as $\mathcal{I}_g$), and $\Omega_g$ is the diagonal matrix of working weights, which are ${\mu_g}/(1+\phi_g\mu_g)$ for the NB model.

\subsection{Moderated and trended dispersion estimates}
The adjusted profile likelihood (APL) introduced by Cox and Reid \cite{Cox1987} has shown good performance for dispersion estimation in the context of genome-scale count data \cite{McCarthy2012,Robinson2008}. The $APL_g$ is a likelihood in terms of $\phi_g$, penalized for the estimation of the regression parameters, $\beta_g$, as follows:
\begin{equation}
\label{eq:est_phi}
APL_g(\phi_g) = \ell (\phi_g; \mathbf{y}_g, \hat{\beta_g}) - \frac{1}{2}\log |\mathcal{I}_g|,
\end{equation}
where $\mathbf{y}_g$ is the vector of counts for gene $g$, $\hat{\beta_g}$ is the estimated coefficient vector, $\ell()$ is the log-likelihood function, $\mathcal{I}_g$ is the Fisher information matrix and $|.|$ is the determinant.  The early strategy to accomplish moderation for the dispersion was by squeezing the tagwise dispersion toward a common dispersion that is estimated over all features \cite{Robinson2007}.  This weighted likelihood approach involves maximizing a linear weighting of the individual likelihood and the common (averaged) likelihood, the two terms, respectively, in:

\begin{equation}
\label{eq:com_apl} 
\arg \max \left\{ APL_g(\phi_g) + \alpha \cdot \frac{1}{I}\sum_{k=1}^{I} APL_k(\phi_g) \right\},
\end{equation}
where $\alpha$ is a suitably chosen weight.

A slight variation on this, which is now commonly applied after experience in many datasets showing a dispersion-mean relationship, is to shrink towards a dispersion estimated from features with similar average expression level \cite{McCarthy2012}.  This so-called trended dispersion is constructed using local shared log-likelihood for feature $g$ (more precisely, a smooth fit to common dispersions that are calculated in bins of averaged counts per million) and its neighbouring features in terms of expression strength.  Specifically, individual tagwise estimates for each feature can be estimated by maximizing a linearly weighted function between individual dispersion and local shared dispersion:

\begin{equation}
\label{eq:tag_apl}
\hat{\phi}_g = \arg \max \left\{ APL_g(\phi_g) + \gamma \cdot APL^S_g(\phi_g) \right\},
\end{equation}
where $\hat{\phi}_g$ is moderated tagwise dispersion, $\gamma$ is the prior degrees of freedom afforded to the shared likelihood and:

\begin{equation}
\label{eq:tag_apl_shared}
APL^S_g(\phi_g) = \frac{1}{|C|} \sum_{k \in C} APL_k(\phi_g) ,
\end{equation}
where the set $C$ represents features that are close to feature $g$ in average log counts per million.





\subsection{A robust negative binomial GLM}

Our approach to induce robustness is to attach a weight to each observation; observations that deviate strongly from the model fit are given lower weight.  In particular, Pearson residuals from the current fit are sent through a weight function, which gets passed to the next iteration of estimation. A flow chart of our robust method given the components discussed below is presented in Figure \ref{flow}.  The dispersion estimation machinery (i.e., trended APL) also receives the same observation weight, so that the influence of outliers is dampened on both the regression and dispersion estimates.
The Pearson residual of an observed count $y_{gi}$ from the NB GLM fit can be calculated as:
\begin{equation}
\label{eq:residual}
r_{gi} =  \frac{y_{gi} - \hat{\mu}_{gi}}{\sqrt{\hat{\mu}_{gi}(1 + \hat{\phi}_g\hat{\mu}_{gi})}}
\end{equation}
where $\hat{\mu}_{gi}$ is the fitted value (from $\hat{\beta}$) and $\hat{\phi}_g$ is the moderated dispersion estimate. The Pearson residual is converted to weights using, for example, the Huber function:

\begin{equation} 
\label{eq:weight}
w_{gi} = w(r_{gi}) =\begin{cases}
                          \frac{k}{abs(r_{gi})}, &\text{for abs}(r_{gi})>k\\
                          1,                           &\text{for abs}(r_{gi}) \leq k
                      \end{cases}  
\end{equation}
where $k$ represents a tuning constant for Huber estimator and usually set to $1.345$ in normally-distributed settings to achieve 95\% efficiency \cite{Fox2002}. This weight, $w_{gi}$, gets used in the next iteration of GLM fitting; the IRLS Equation becomes:
\begin{equation}
\beta_g^\text{W-new} = {\beta_g}^\text{W-old} + (X^T [W_g \Omega_g] X)^{-1} X^T [W_g] z_g
\end{equation}
where $W_g$ the diagonal matrix of observation weights for feature $g$. The Fisher information matrix with observation weight becomes  $\mathcal{I}^W_g = X^T [W_g\Omega_g] X$.  In this approach, the APL for dispersion $\phi_g$ with observation weights can be written as:
\begin{equation}
\label{eq:aplw}
APL^W_g(\phi_g) = \ell^W (\phi_g; \mathbf{y}_g, \hat{\beta_g}) - \frac{1}{2}\log |\mathcal{I}^W_g|, 
\end{equation}
where  $\ell^W(.) \equiv \sum_{i} w_{gi} l(.)$ is the weighted log-likelihood function and $\mathcal{I}^W_g$ is the Fisher information matrix with observation weights. Then, using these dispersion estimates, the regression parameters are estimated, again using the observation weights. 


For users of \tool{edgeR}, only a small change in the standard pipeline is required.


\begin{figure*}[ht]                 
\centering
\includegraphics[width=\textwidth]{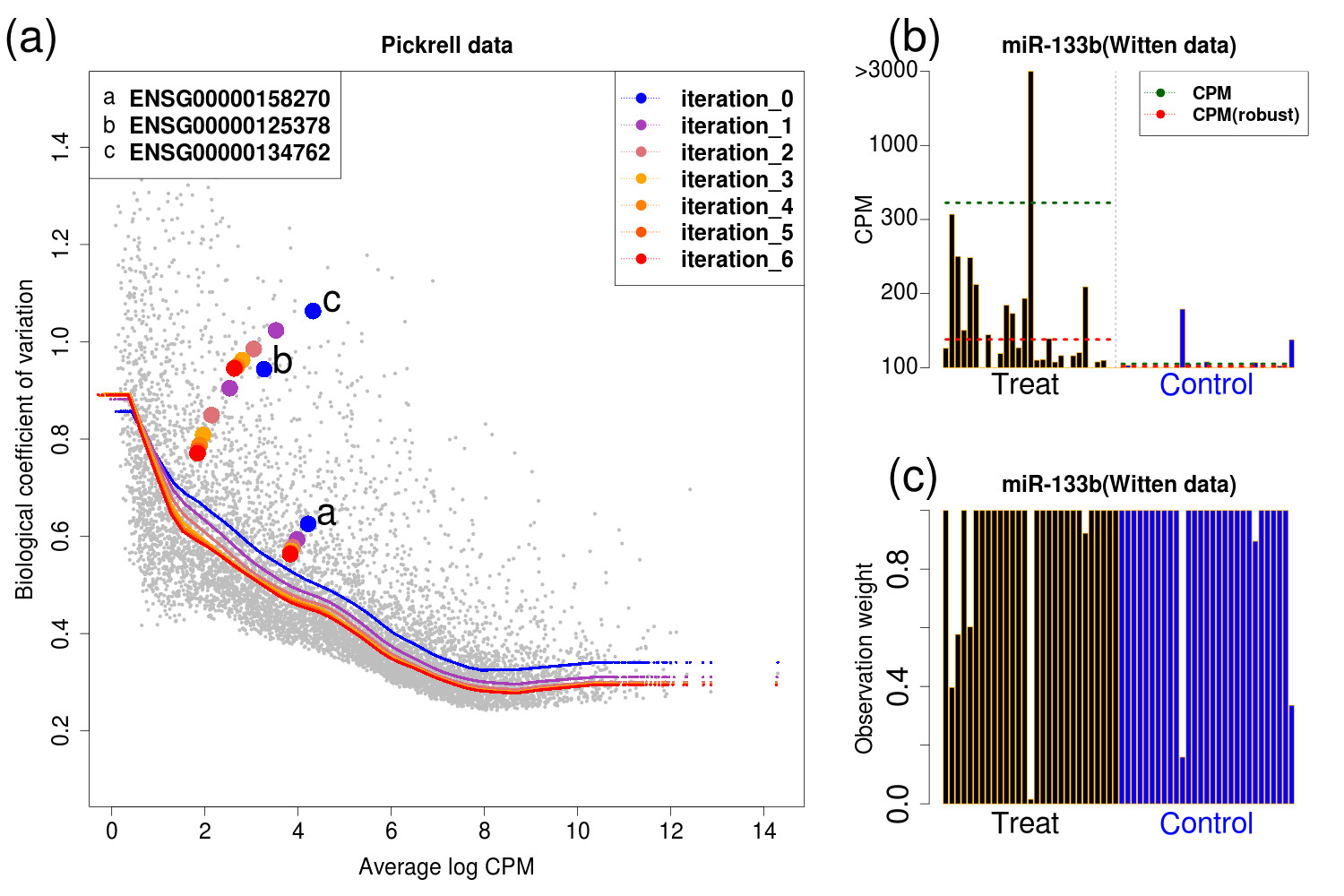}
\caption{(a) For the random 5 versus 5 split of the Pickrell data \cite{Montgomery2010} shown in Figure \ref{fig:illustration}, the trajectories of overall trended dispersion and for the 3 individual genes are shown over 6 iterations of the \tool{edgeR-robust} reweighted estimation scheme. (b) a barplot of miR-133b expression from Witten et al. \cite{Witten2010}, including an observation with very high count. (c) weights for miR-133b after 6 iterations of the reestimation from \tool{edgeR-robust}.  Dashed lines in panel (b) shown the group-wise counts-per-million (CPM) before and after weighting.} 
\label{fig:after_robust}
\end{figure*}

\subsection{A simulation framework with parameters based on the joint distribution of mean and dispersion estimates from RNA-seq data}

We built a simulation framework that aims to accurately reflect the reality of RNA sequencing data. In order to evaluate the performance of our robust method and other methods across a variety of reasonable conditions, we created several options:
\begin{enumerate}
\item \Rcode{nTags}: total number of features
\item \Rcode{group}: factor containing the experimental conditions
\item \Rcode{pDiff}: proportion of DE features
\item \Rcode{foldDiff}: relative expression level of truly DE features
\item \Rcode{pUp}: proportion of features with an increase in expression
\item \Rcode{dataset}: dataset to take model parameters from
\item \Rcode{pOutlier}: proportion of outliers to introduce
\item \Rcode{outlierMech}: outlier generation mechanism to use
\end{enumerate}
We generate true NB model parameters, $\mu$ and $\phi$, using the joint distribution of estimates, $\hat{\mu}$ and $\hat{\phi}$, from a real datasets, such as the published count tables at \tool{ReCount} \cite{Frazee2011}: Pickrell \cite{Montgomery2010}, Cheung et al.\cite{Cheung2010, Witten2010}.  In particular, the joint distribution preserves the dispersion-mean trend, which can vary from dataset to dataset.  After the removal of extremely high dispersions and low means (analogous to typical recommended filters), the derived-from-real-data parameters are used to simulate the counts, from a NB distribution and optionally with true DE. 

To test robustness, we add outliers to the simulated counts. Outliers are large values and can be produced by two different mechanisms (\Rcode{outlierMech}): first, counts are multiplied by a random factor between $2$ and $10$, as employed by Soneson and Delorenzi \cite{Soneson2013}, and includes both the ``single" (S) and ``random" (R) method. In S, a gene is chosen at some probability to have a single outlier randomly added. In R, each observation can become an outlier with some probability.  The second mechanism, called ``model" (M), each observation can be become an outlier with some probability and if so, is sampled from a second NB distribution with large $\mu$; R and M methods induce the same overall outlier rate.

Recently, van de Wiel et al. modeled genome-scale count data as zero-inflated negative binomial model (ZINB), which seemed to explain some of the dispersion-mean relationship \cite{VanDeWiel2013}.  We have not considered simulations from ZINB distributions, since they do not appear to explain all of the observed dispersion-mean relationship observed in the datasets that we tested (see Supplementary Figure 1).
 



\subsection{Methods compared}
We evaluated and compared several methods for DE analysis, including \tool{edgeR}, \tool{edgeR-robust}, \tool{limma-voom}, \tool{DESeq-pool}, \tool{DESeq-glm}, \tool{DESeq2}, \tool{baySeq}, \tool{SAMseq} \cite{Li2011a}, \tool{EBSeq} \cite{Leng2013} and \tool{ShrinkBayes}; the R-based performance evaluation system that we developed allows arbitrary additions (assuming they are implemented in R). \tool{limma-voom} is an extension to DE analysis of RNA-seq count data from \tool{limma} \cite{Law}; it transforms the count data with special treatment given to fitting the mean-variance relationship. \tool{DESeq} is tested as two separate methods: \tool{DESeq-pool} is the default setting method to estimate the empirical dispersion from all the conditions with replicates; \tool{DESeq-glm} fits models according to a design matrix and estimates dispersion by maximizing APL.  \tool{edgeR}, \tool{DESeq} and \tool{DESeq2} differ in how the dispersion is estimated: \tool{edgeR} moderates dispersion towards a trended estimate \cite{McCarthy2012}, \tool{edgeR-robust} expands this with observation weights, \tool{DESeq} takes the maximum of a fitted trend of dispersion or the individual feature-wise dispersion estimate \cite{Anders2010}. At time of writing, the details of \tool{DESeq2} were not yet published, but from the documentation, the approach offers a zero-mean normal prior on the regression parameters of interest with an empirical variance estimated from data; for outlier protection, a Cook's distance is calculated and those features with an extreme value are not promoted to formal statistical testing. The default normalization method is also different among \tool{edgeR}, \tool{DESeq} and \tool{DESeq2}. \tool{edgeR} uses trimmed-mean-of- M-values (TMM) \cite{Robinson2010}, while \tool{DESeq} and \tool{DESeq2} use a relative-log-expression approach. \tool{SAMseq}, a non-parametric method, employs Wilcoxon rank-sum statistics to estimate false discovery rate (FDR) through sample permutations. 

\tool{baySeq}, \tool{EBSeq} and \tool{ShrinkBayes} use Bayesian inference. \tool{baySeq} employs the NB model and assumes that samples can be classified as different groups by their treatment conditions; samples within the same group should follow the same distribution and share parameters. Using an empirical Bayes approach, \tool{baySeq} estimates the posterior probability of the null state. \tool{ShrinkBayes} introduces the ZINB and performs inference using integrated nested Laplace approximations (INLA) \cite{Rue2009, 2012arXiv1210.0333M} and provides Bayesian FDR and local false discovery rate (lfdr) \cite{Efron2001} estimates. Since the computational cost of \tool{ShrinkBayes} is high, some comparisons are skipped. \tool{EBSeq} is similar to \tool{baySeq}, providing posterior probability of DE, as well as EE (equally expressed), based on a parametric mixture model. Compared with other methods tested here, \tool{EBSeq} can also detect DE isoforms in EE features, yet this is not our primary question here.

Notably, new methods, or variations of existing ones can be easily added to our comparison framework, simply by providing a wrapper to an \tool{R} function that contains the correct inputs (count table, grouping variable) and outputs (P-values).  See Supplementary web site for details.

\subsection{Comparison metrics}
To test the performance of each DE method (in the presence of outliers), we employ several standard metrics and plots: false discovery (FD) plots, receiver operating characteristic (ROC) curves, partial ROC curves and power curves. Power (TP) curves and (partial) ROC curves (i.e., up to a certain false positive rate) evaluate the ability to distinguish, through statistical evidence, DE and non-DE. FD procedures gauge the control of the expected proportion of incorrectly rejected null hypotheses \cite{Benjamini1995}.  Another useful plot is the relationship between TP rate and achieved false discovery rate across multiple thresholds.





\subsection{An open graphical tool and R code for re-analysis: evaluating DE analysis methods} 

One disadvantage of current method comparisons (e.g., \cite{Soneson2013, 2013arXiv1301.5277R}) and those that accompany every new method published, is that they are a snapshot in time.  If new  methods come along, the developer must demonstrate that their method is better, by some metric.  This task is important but somewhat repetitive, since many of the same metrics, plots and simulation models are (re-)implemented.  We endeavored to create a system for performing standardized simulation-based testing.

In addition, all analyses presented in this paper are freely available from our website. Moreover, our simulation and evaluation framework is made available as a web-sourceable script that consists of 3 modules:  simulation, evaluation (running of the software packages) and metric computation.  Each module can be extended, using simple wrapper functions to existing R-based code, ensuring that our comparison results are reproducible, extensible and relatively easy for the user to track exactly what code segments (and versions) were run.

In addition to R code, we make available a web-based \tool{shiny} ``app'' that can be used to look at simulation results across a wide number of conditions \cite{Rshiny2013}.  Since there are often too many methods to be easily displayed together, our app gives users the ability to present results for a user-selected subset of methods; the results update automatically as the user selects different simulation settings.



\subsection{Functional category analysis for outliers} 
To explore potential biological or technical factors that may manifest as outliers, we performed hypergeometric-based functional category analyses on the set of genes with weights less than some cutoff (here, set to 1) on a per-sample basis.  Our goal with such an analysis is to identify
possible biological or technical factors that affect a subset of genes for a particular experimental unit.  In some cases, this may shed light on why the expression levels of some genes for a given sample are very different than that of their replicates.  Furthermore, we can investigate whether the downweighting is driven by technical factors.  As a positive control for this, we compared the observed weights to the sample-specific GC content effects observed in the Pickrell dataset \cite{Hansen2012, Montgomery2010}
 
\section{RESULTS}

\subsection{edgeR-robust dampens the effect of outliers}

To highlight how \tool{edgeR-robust} dampens the influence of outliers, we return to the dataset shown in Figure 1. Figure~\ref{fig:after_robust}(a) shows the trajectories for the 3 outliers in terms of their average log-CPM and dispersion estimates and how the dispersion-mean trend changes over 6 iterations of the \tool{edgeR-robust} reweighted estimation scheme.  As expected, the outliers appear ``extreme'' according to the model, as also reflected by their residuals.  Extreme residuals are then down weighted, iteratively, and both the dispersion and average log-CPM estimates are altered (Figure~\ref{fig:after_robust}a).  In particular, we notice large changes to both the average log-CPM (or equally, to the regression parameter estimates) and they become in better accordance with the other features in the dataset.  Notably, Figure~\ref{fig:after_robust}(a) highlights a global drop in dispersion-mean trend after the iterative robust estimation, which suggests that outliers present in sufficient frequency may have a global effect on the statistical detection of DE within a dataset.  Thus, we speculate that gains in statistical power (see Sections below) may be achieved in part by this global drop in trended dispersion.

In their manuscript, Li and Tibshirani \cite{Li2011a} show some extreme examples of outliers affecting differential count analysis of miRNA-seq data (in particular, see their Figure 2).  Figure \ref{fig:after_robust}(b) shows one of those examples, mir-133b, and highlights the estimated mean CPM by group, before and after downweighting; the observation weights after 6 iterations are shown in Figure \ref{fig:after_robust}(c).  Notably, for this example, there still exists strong evidence for differential expression, even after careful reassessment of the outlying observations.

Supplementary Table~\textbf{1} shows the full details of these analyses, before and after reweighting.




\begin{figure*}[ht]                 
\centering
\includegraphics[width=\textwidth]{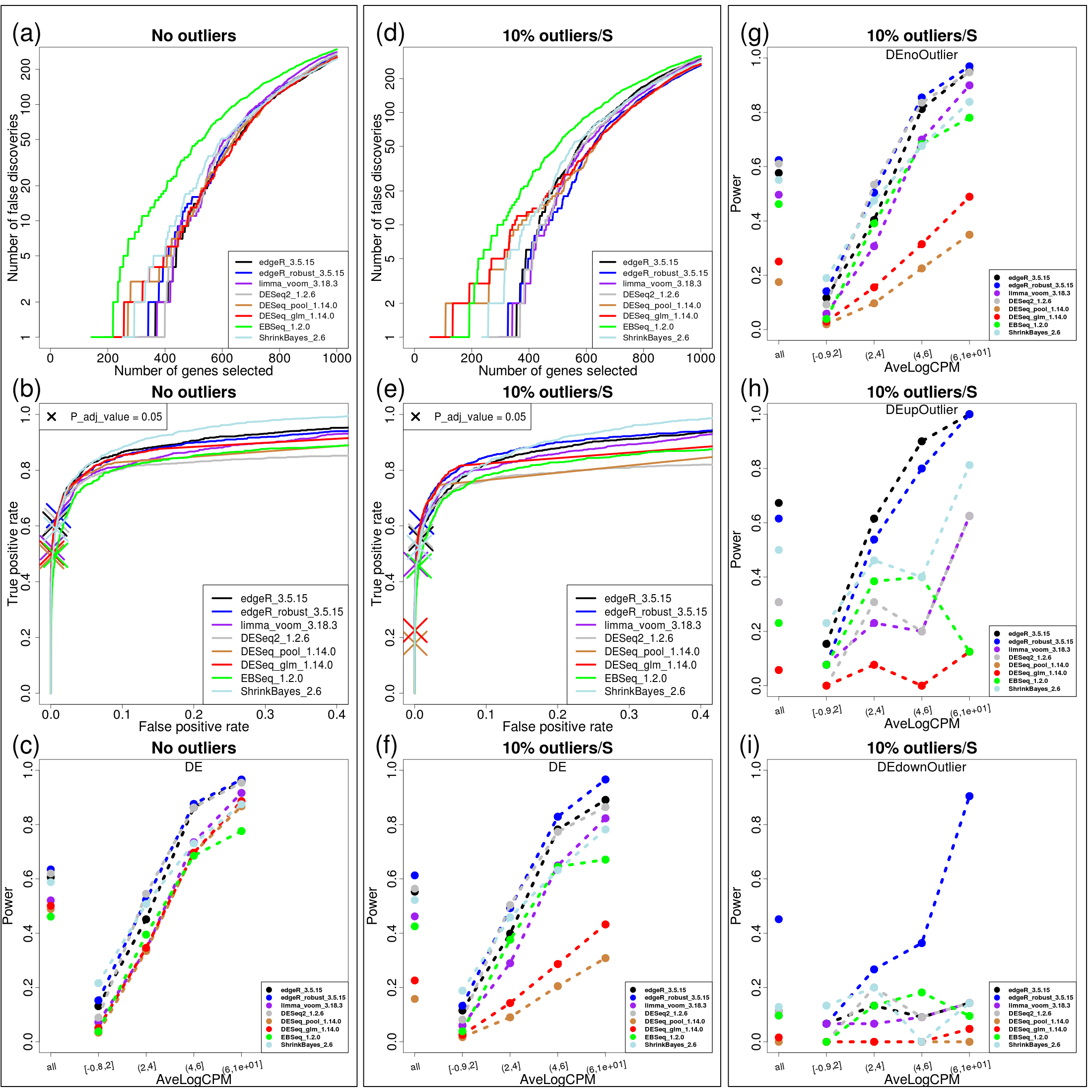}
\caption{(a), (b) and (c) present FD, partial ROC (up to FP rate of 40\%) and power plots (at each methods' 5\% FDR) across several tested methods for datasets with no introduced outliers;  (d), (e) and (f) show corresponding plots with datasets containing 10\% outliers (i.e., 10\% of genes have a single outlier) using ``S" method.  (g), (h) and (i) split the results from panel (f) into three categories: features without outliers (g); outliers in the higher expression group (h); outliers in the lower expression group (i).  All power results are shown as overall (single dot on left of plot) and split across five equally-sized average-log-CPM groups.  The X on panels (b) and (e) highlights the achieved power (TP) according to each method's 5\% FDR cutoff.  Note that while panel (g) presents the situation with no outliers, there are outliers present in other features within the dataset and is therefore different from panel (c).} 
\label{sim_data}
\end{figure*}

\subsection{Simulation reflects real data}

To test the method on a wide range of simulated settings, we first generate count data from a model that reflects real data as well as possible.  As described in the Methods, we choose to take the joint distribution of estimated log-CPM and dispersion for a large dataset, after applying no moderation to the parameter estimates, as the basis for the parameter settings and we use library sizes that mimic those from typical datasets.  For example, the Pickrell dataset \cite{Montgomery2010} consists of more than 50 replicates, which should represent a reasonably accurate reflection of the range of abundances observed as well as, in particular, the dispersion-mean relationship.  We generate all data from the NB model and introduce outliers by various mechanisms (see Methods). Supplementary Figure 2 shows the dispersion-mean trend for the Pickrell dataset (Panel a) and an example simulated dataset based on the estimated parameters (Panel b), respectively, as well as the marginal distributions of both log-CPMs and dispersion.  The framework for these simulations (see Methods) is designed to take an initial dataset that seeds the simulation parameters, so datasets spanning the range of biological variation could easily be tested.

\begin{figure*}[ht]               
\centering
\includegraphics[width=\textwidth]{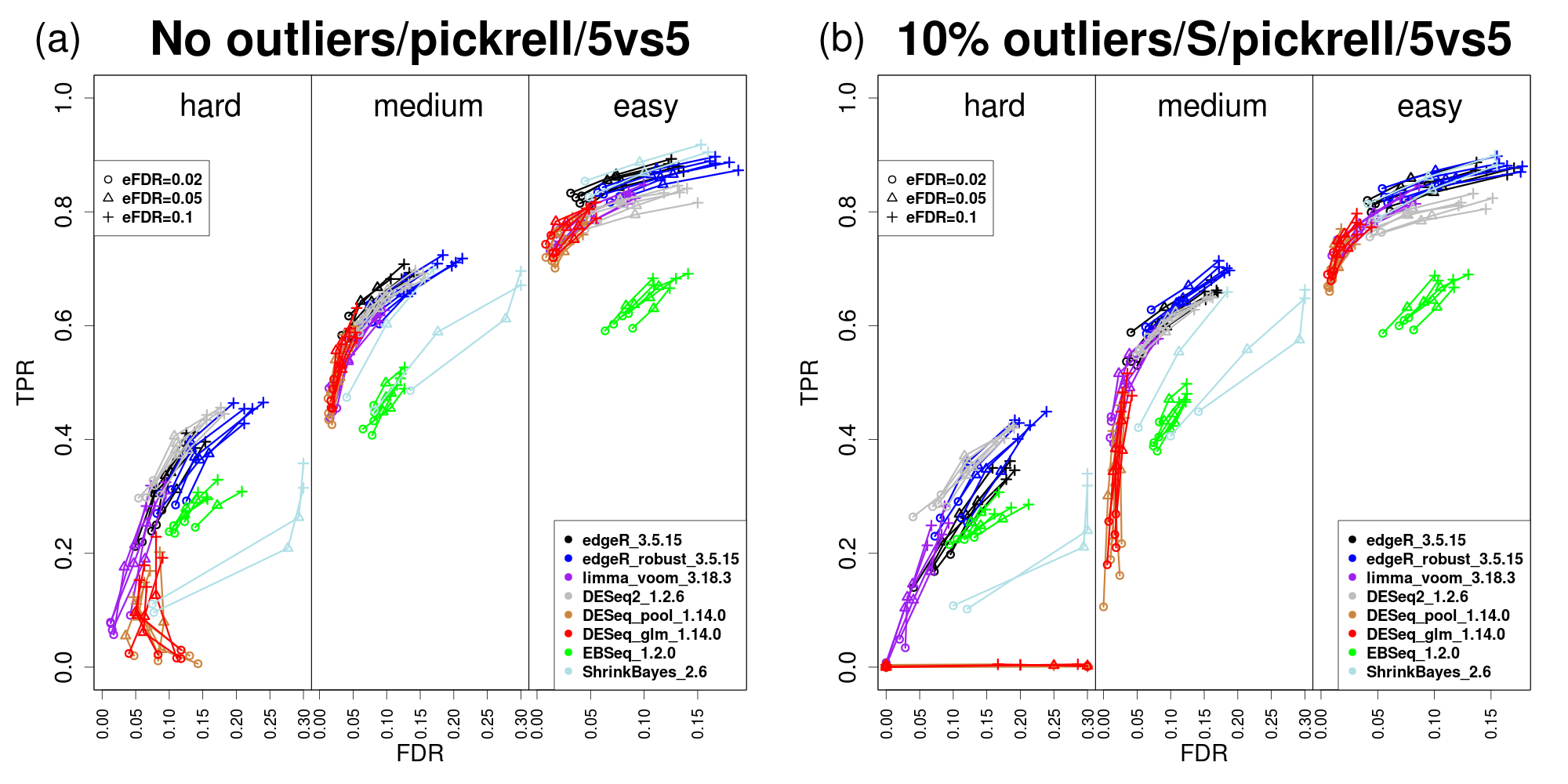}
\caption{Power-to-achieved-FDR across hard (\Rcode{foldDiff} $\in [2,2.2]$), medium (\Rcode{foldDiff} $\in [3,3.3]$) and easy (\Rcode{foldDiff} $\in [6,6.6]$) simulation settings.  (a) No outliers; (b) 10\% outliers.  Y-axis shows TP rate and X-axis shows.  5 simulations are shown for each method and each setting.  Points are taken according to each method's FDR cutoffs at $0.02$, $0.05$ and $0.1$.} 
\label{sim_dataF}
\end{figure*}

\subsection{Standard metrics across various methods for various simulation settings}

Next, we present a representative simulation and performance results under a single ``reasonable'' setting of the parameters.  We sampled NB model parameters $\mu$ and $\phi$ from the joint distribution of estimates from the Pickrell data \cite{Montgomery2010} (\Rcode{dataset}); we filtered out the top $10$ percent of the extreme dispersion values; $10000$ features are generated (\Rcode{nTags}), with a 5 versus 5 two-group comparison (\Rcode{group}); $10\%$ of them are defined as DE genes (\Rcode{pDiff=.1}), symmetrically (\Rcode{pUp=.5}) with fold difference $3$ (\Rcode{foldDiff=3}); outliers are introduced to $10\%$ of the features (\Rcode{pOutlier=.1}) using the ``simple'' outlier generation mechanism (\Rcode{outlierMech="S"}); outliers are randomly distributed amongst all genes; further details are described in Methods. Original simulated counts and the counts with outliers introduced are separately recorded.

Figure \ref{sim_data} shows the set of standard metrics:  panels (a)-(c) and (d)-(f) show false discovery plots, ROC curves and power numbers, respectively, for the original and original-with-outliers datasets under the setting of simulation parameters discussed above.  Overall, the introduction of outliers results in more false positives (Figure~\ref{sim_data}a versus \ref{sim_data}d) and/or less true positives at the same false positive rate (Figure~\ref{sim_data}b versus \ref{sim_data}e).  In the absence of outliers, all methods exhibit similar patterns of false discovery rates, with the Bayesian methods, \tool{ShrinkBayes} and \tool{EBSeq} having a slightly higher rates.  Similarly, in terms of separating the truly DE from non-DE features using a P-value (or P-value-like score in the case of Bayesian methods), all methods are very close in performance.  When outliers are introduced, \tool{edgeR-robust} shows some advantages over \tool{edgeR}, but many other methods are close.  In terms of statistical power, all methods drop in overall power with the introduction of outliers (Figure~\ref{sim_data}c versus \ref{sim_data}f), but \tool{edgeR-robust} drops by the smallest amount. While \tool{DESeq} maintains a good ranking of P-values, it becomes very conservative due to the maximum-of-trend-and-individual dispersion policy; in this respect, presence of outliers affect the whole dataset (see Supplementary Figure 3).

Since the direction of differential expression and the outlier introduction are applied at random, we can further split the DE features according to the position of the outlier relative to the direction of change in abundance (Figure~\ref{sim_data}g-i; ``DEupOutlier'' represents the situation where the outlier is added to the higher expressed group; ``DEdownOutlier'' represents those features where the outlier was added to the lower expressed condition; ``DEnoOutlier'' represent DE features with no introduced outlier).  Notably, \tool{edgeR} shows the highest power in the ``DEupOutlier'' setting, but this is artificial since the introduction of the outliers actually helps detection in this situation.  The ``DEdownOutlier'' is the situation where \tool{edgeR-robust} comes to the forefront, as expected, given that outliers strongly eliminate the differential expression in this setting.  In the absence of outliers, \tool{edgeR-robust} still remains a strong competitor, closely followed by \tool{DESeq2}, \tool{ShrinkBayes} and \tool{edgeR}.

It is also interesting to consider how well the robust observation-weight-based method and \tool{DESeq2}'s Cook's distance policy perform in simply identifying the simulated true outliers.  Supplementary Figure 4 shows an ROC curve depicting how well the observation weights separate outliers from non-outliers.  Similarly, the default setting of \tool{DESeq2} leads to a similar tradeoff between false positives (here, falsely detected as an outlier) and false negatives (failing to identify an outlier).  Notably, the \tool{edgeR-robust} strategy smoothly identifies outliers and downweights them according to the magnitude of discordance, instead of setting a hard threshold where statistical tests are no longer conducted.  One byproduct of \tool{DESeq2}'s hard threshold is a loss of power (e.g., Figure \ref{sim_data}panels h and i), since genes with true differential expression are excluded from statistical testing.

The above discussion was in regard to a single dataset under a single set of simulation parameters.  To provide a much wider scope of simulation settings, we created a web-based \tool{shiny} app, that serves up pre-computed results over a range of simulation parameters, including different datasets, sample sizes and so on.  In addition, it allows users to plot results for only the subset of desired methods and metrics from Figure~\ref{sim_data}.  While new methods can only be added to the \tool{shiny} app by us, the same simulations can be easily recreated in a local \tool{R} environment, as described on our website.  In general, the conclusions observed from the broader range of simulation settings are in agreement with those mentioned above (Supplementary Figure 5).  We also tested \tool{DESeq2} when turning off the Cook's distance metric and found Pearson residuals to outperform deviance residuals (Supplementary Figure 6).

\subsection{Across multiple simulations over a range of settings, \tool{edgeR-robust} is somewhat liberal but achieves the best power-to-effective-FDR tradeoff}

To complement the simulation results for individual parameter settings, we endeavoured to create a compact summary of a wider range of simulations and explore another important aspect of these methods: do methods accurately report false discovery rate?  Figure~\ref{sim_dataF} shows a series of $15$ simulations divided into $3$ different blocks based on the degree of difficulty: ``hard" (\Rcode{foldDiff}$\in [2,2.2]$), ``medium'' (\Rcode{foldDiff} $\in [3,3.3]$) and ``easy" (\Rcode{foldDiff}$\in [6,6.6]$), including $5$ simulations within each group to illustrate sampling variability.  For each dataset, lines connect the true positive rates and effective FDRs across three thresholds of the estimated FDR (.02, .05, .1). The rest of the simulation parameters are kept fixed: the NB model parameters originate from the Pickrell dataset \cite{Montgomery2010}, there are $10000$ features, we consider a two group comparison ($5$ versus $5$), $10\%$ of features are DE and each dataset contains $10\% $ ``S" outliers; comparisons for $3$ versus $3$ and $10$ versus $10$ are shown in Supplementary Figure 7.

Overall, there is a broad range of power-to-achieved-FDR tradeoffs.  On the one hand, \tool{EBSeq} appears to be a bit conservative and more so in large samples, but quite liberal in small samples. In general, \tool{DESeq} is conservative and achieves lower power, as reported earlier \cite{Soneson2013}.  Altogether, the collection of methods, such as \tool{voom}, \tool{edgeR}, \tool{edgeR-robust} and \tool{DESeq2} achieve similar power-to-achieved-FDR tradeoffs across the sample sizes, with perhaps a tendency to be more liberal in large sample sizes for \tool{edgeR} and \tool{edgeR-robust}.  As expected and as highlighted above, \tool{edgeR-robust} appears to have advantages in the presence of outliers, with only a minor decrease in power when no outliers are present.  Overall, edgeR achieves the best tradeoff between power at the same achieved FDR, even if the target FDR is not quite met.

\begin{figure*}[ht]               
\centering
\includegraphics[width=\textwidth]{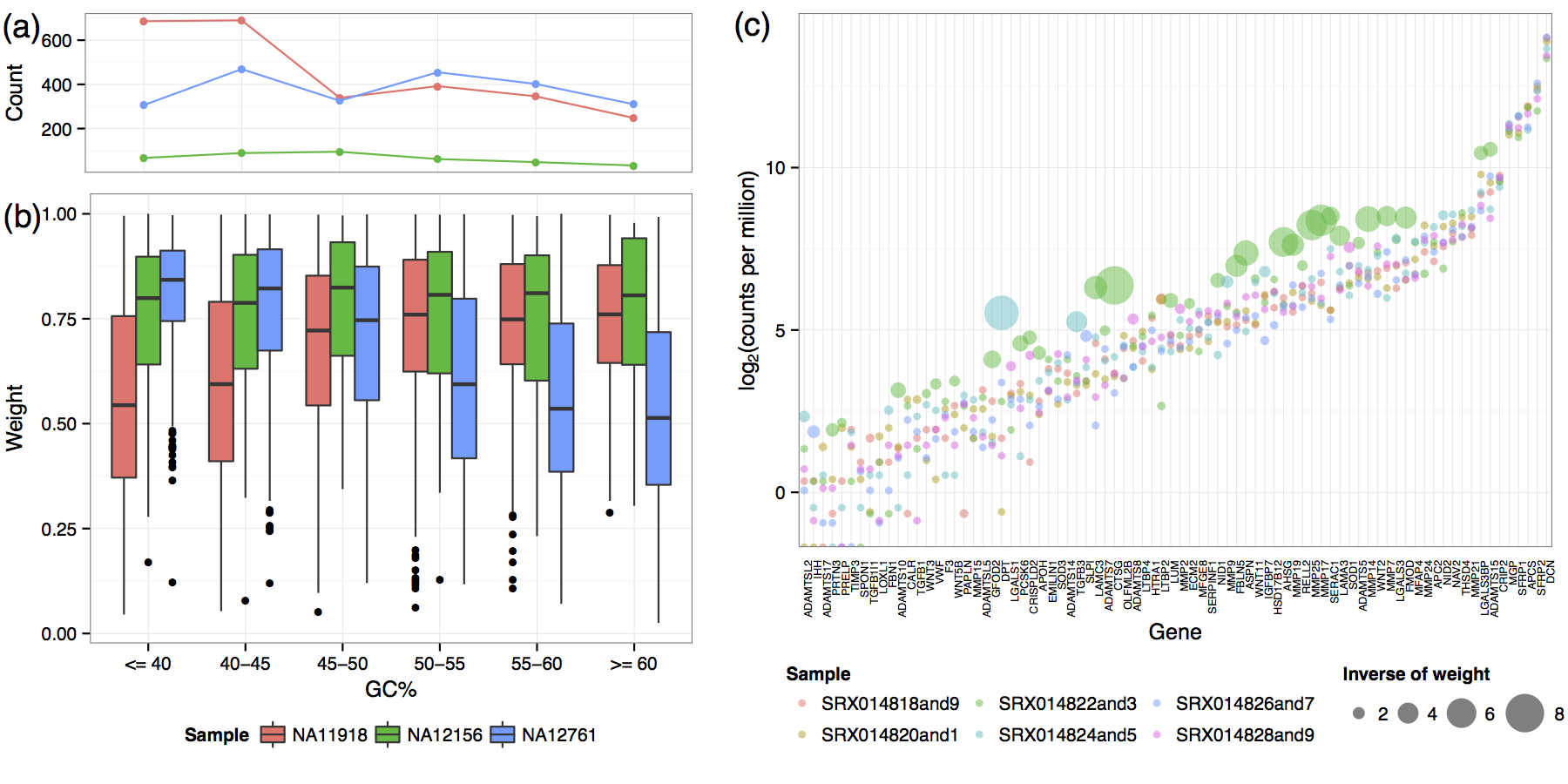}
\caption{Technical ((a) and (b)) and biological (c) sources of outlier genes. The number of down weighted observations (a) and  distribution of outlier weights as a function of the gene GC\% in three samples from the HapMap RNA-Seq data \cite{Montgomery2010} are plotted (b).  Two of the samples shown (NA11918 and NA12761) were shown by Hansen et al to have strong, opposing relationships between GC\% and RPKM.  The third sample (NA12156) had the least number of genes down weighted after applying our robust down weighting procedure.  (c) shows the \(log\)(CPM) and the inverse of the down weighting value for genes in the ``extracellular matrix" gene ontology category, where a value of 1 indicates no down weighting and larger inverse weights indicate stronger down weighting. } 
\label{f:outlier_explore}
\end{figure*}

\subsection{Outliers may originate from technical or biological sources}

While the strategy based on observation weights appears useful for dampening the effect of outliers in differential expression analysis, it may also be of interest to investigate the origin of such outlying observations.  In some cases, we know of technical artifacts that affect the profile of RNA-seq expression data, such as sample-specific GC content biases, as highlighted and mitigated by the analyses of the HapMap consortium as well as in follow-up methodology development (e.g., cqn normalization \cite{Hansen2012}).  In this dataset, there are no experimental conditions to detect differential expression, so we fit an intercept-only model, using the iterative robust estimation scheme.  Not surprisingly, we first observe that the two samples highlighted by Hansen et al. also exhibit a relatively higher number of down weighted observations (Figure \ref{f:outlier_explore}a).  As expected, the degree of downweighting is strongly related to the GC content of the cDNA sequences of the genes involved (Figure \ref{f:outlier_explore}b).

In an unrelated dataset from Blekhman \cite{Blekhman2010} comparing expression in human male and female livers, we observe that the most significantly overrepresented functional categories were associated with a signal sample (SRX014822and3, green circles in Figure \ref{f:outlier_explore}c).  These include several categories involving the extracellular matrix, as well as collagen catabolism and plasma membrane.  We show the third most overrepresented category, ``extracellular matrix" (Figure \ref{f:outlier_explore}c) because the size of this category allows individual genes to be visualised (further details are given in Supplementary Table 2).  Although we cannot confirm the exact cause of the overrepresented gene ontology categories, we note that accumulation of collagen and excessive production of extracellular matrix proteins are associated with the development of liver fibrosis (e.g. \cite{Herbst, Asselah2005}), and we suggest that analyses such as these may assist biologists in identifying the source of outliers in gene expression.



\section{DISCUSSION}

Various method developers have shown that statistical methods for the task of discerning differential expression from RNA-seq data represented as counts can be sensitive to outlying observations.  In this report, we have studied in detail the effects of outliers on various approaches and developed a new method based on observation weights that can detect and dampen the effect of outliers.  In fact, it requires a delicate tradeoff between maintaining high power while at the same time achieving a decent resistance to the presence of outliers.  In particular, it is difficult to know exactly what an outlier is and where the line should be drawn to identify it as such.  In this respect, we take a ``smooth'' approach of dampening their effects, when there is evidence to support departure from the model.  We have also explored the origin of such outliers and in some cases, we may be able to identify either a technical or biological effect to explain them.  Our robust approach follows the strategy of classical robustness methods that are commonly applied to the linear regression problem.  In our approach, we adopted the calculation of the residuals and observation weights to the specifics of the flexible dispersion estimation and standard GLM regression estimation of the negative binomial model.

As mentioned above, one reason the \tool{edgeR} is sensitive to outlying observations is that the dispersion estimate used in the downstream inference is pulled towards the dispersion-mean trend, which may underestimate the variability.  Therefore, another way to dampen the effect of outliers is to decrease the degree of moderation toward the dispersion-mean trend.  Although we have not studied it here, there is again a delicate tradeoff between the degree of moderation to use and the average inference performance; it still remains an open question as to how exactly to set this value for a given dataset.
 
Though motivated and tested on real datasets, we employed simulations to explore the broad range of possible settings and developed a comprehensive system for such evaluations.  Our strategy to mimick real datasets is to take the joint distribution of mean and dispersion estimates from a large dataset as the basis for parameters to sample from.  From such a dataset, outliers and differential expression at a specified level can be readily introduced.  In fact, because these are estimates and not true values, we expect the sampled dispersion to potentially exhibit more variation than observed in a real dataset.  In terms of evaluating the different methods across the spectrum of simulation settings, it is important to consider it from all points of view: false discoveries amongst the list of top called features, the ability to separate the truly differential from non-differential (i.e., ranking by statistical evidence), the statistical power at thresholds that are typically used in practice and the degree to which methods achieve their purported false discovery rates.

Overall, the observation weight robust method performs well and achieves the goal of suffering only minimal loss of power, while maintaing resistance to introduced outliers.  We have investigated the outlier policy in other packages and highlight that smoothly downweighting outlying observations appears preferable.  In \tool{DESeq}, a hard line against outliers is taken by using the maximum of a dispersion-mean trend and the individual estimate; with the addition of outliers, this has a global effect of increasing the variance to all features and gives a resulting loss of power.  In \tool{DESeq2}, a Cook's distance metric is used to remove features with outliers entirely from further consideration; in this case, features that have outliers and differential expression are excluded, again resulting in loss of power.

With the simulation system that we have created, we can now make a call to the community of both developers and users to check the effect of various settings.  All that is required to test a new method and compare it against existing methods is to write a wrapper function with the correct inputs and outputs.  In addition, if the exact simulation settings that we use in this report are not adequate, we can easily extend this framework into an open testing system that allows additional variations on the sampling model, perhaps including additional distributions or constructed truths, such as plasmodes \cite{10.3389/fgene.2013.00178}.

The current \tool{edgeR} framework does not always achieve its false discovery rate target.  However, even if it were forced to be more conservative, it still achieves power as good or better than existing approaches across the simulation settings that we have tested, even with the addition of observation weights.  The exact source of the liberality is beyond the scope of the current investigation, but there may be room for improvement, such as borrowing ideas from small sample asymptotic approximations \cite{DiYanming2013}.

\section{CONCLUSION}

We developed an approach to dampen the effect of outliers on count-based differential expression analyses.  Overall, the method appears to achieve the desired ``efficiency'': a resistance to outliers while maintaining high power.  We provided an implementation for the \tool{edgeR} Bioconductor package, but the reweighting idea could easily be adopted to other packages.  In addition, we developed an extensible simulation system (at the count table level) that readily performs the simulations based on an existing dataset and provides the infrastructure for producing the standard battery of evaluations.  In particular, this allows new methods or variations (e.g., alternative settings) of existing packages to be quickly explored.  Instead of preparing a large number of Supplementary Figures, we provide an interactive web-based \tool{shiny} ``app'' to display simulation results across a broad range of simulation settings.

\section{ACKNOWLEDGEMENTS}

The authors wish to thank all members of the Robinson laboratory for helpful discussions and in particular, Olga Nikolayeva, Gosia Nowicka, Katarina Matthes and Charity Law for careful reading of an earlier version of the manuscript; we also thank members of the Baudis and von Mering groups for useful feedback.  We thank Gordon Smyth and Aaron Lun for aspects of the edgeR implementation.  We wish to acknowledge funding from an SNSF Project Grant (143883) and the European Commission through the 7th Framework Collaborative Project RADIANT (Grant Agreement Number: 305626).

\subsubsection{Conflict of interest statement.} None declared.

\small
\bibliographystyle{plain}
\bibliography{myrefs}

\end{document}